# A new approach to probabilistic population forecasting with an application to Estonia

**David A. Swanson**
**University of California Riverside**
**Center for Studies in Demography and Ecology, University of Washington**
dswanson@ucr.edu

**Jeff Tayman**
**Tayman Demographics**
**San Diego, California**
jtayman@san.rr.com

**Abstract.**

This paper shows how measures of uncertainty can be applied to existing population forecasts using Estonia as a case study. The measures of forecast uncertainty are relatively easy to calculate and meet several important criteria used by demographers who routinely generate population forecasts. This paper applies the uncertainty measures to a population forecast based on the Cohort-Component Method, which links the probabilistic world forecast uncertainty to demographic theory, an important consideration in developing accurate forecasts. We applied this approach to world population projections and compared the results to the Bayesian-based probabilistic world forecast produced by the United Nations, which we found to be similar but with more uncertainty than found in the latter. We did a similar comparison in regard to sub-national probabilistic forecasts and found our results to be similar with Bayesian-based uncertainty measures. These results suggest that the probability forecasts produced using our approach for Estonia are consistent with knowledge about forecast uncertainty. We conclude that this new method appears to be well-suited for developing probabilistic world, national, and sub-national population forecasts.

## 1. Introduction

In a seminal paper, Alkema *et al.* (2015) describe a Bayesian approach that links probabilistic uncertainty to a world population forecast based on the Cohort-Component Method (CCM). It proceeds by assembling a large sample of future trajectories for an outcome such as the total population size. The point projection in a given year is the median outcome of the sample trajectories. Other percentiles are used to construct prediction intervals (Alkema *et al*., 2015: 2). More details on this approach are found in Raftery, Alkema, & Gerland (2014), and a general overview of probabilistic population forecasting can be found in Raftery & Ševčíková (2023).

Because the Bayesian approach described by Alkema *et al*. (2015) is based on the CCM, its measures of uncertainty are linked to the "fundamental equation," whereby a population at a given point in time, $P_{t+k}$, is equal to the population at an earlier point in time, $P_t$, to which is added the births and in-migrants that occur between time t and time t+k and to which is subtracted the deaths and out-migrants that occur during this same time period (Baker *et al*., 2017: 251–252). The fundamental equation is the cornerstone of demographic theory and is the foundation upon which the CCM rests (Baker *et al.*, 2017; Burch, 2018). A probabilistic approach to population forecasting based on this theoretical foundation yields benefits not found in methods lacking this foundation (e.g., Burch, 2018; Land, 1986). This observation is also consistent with one made by Swanson *et al*. (2023), who argue that a given population forecasting method's strengths and weaknesses largely stem from four sources: (1) Its correspondence to the dynamics by which a population moves forward in time; (2) the information available relevant to these dynamics; (3) the time and resources available to assemble relevant information and generate a forecast; and (4) the information needed from the forecast.

The Bayes CCM approach comes with strengths. However, it also comes with weaknesses. Goodwin (2015) finds Bayesian inference difficult, effortful, opaque, and counter-intuitive. Along with the weaknesses described by Goodwin (2015) are implied ones, including being not easy to apply or explain and having a low face validity and high production costs in that a Bayes CCM approach is very data- and analytically intensive.

**A New Approach**

We describe an approach for constructing uncertainty measures that is relatively simple and linked directly to the CCM approach. Importantly, unlike Bayesian inference, we believe it is likely to meet important evaluation criteria used by demographers who produce population projections (Smith, Tayman, and Swanson, 2013: 301- 322): Low production costs (particularly staff time); easy to apply and easy to explain; a high level of face validity; and intuitive. In describing this new approach, we use national population projection for Estonia found at the U.S. Census Bureau's International Data Base (U.S. Census Bureau, 2020). Before showing these results, we also employ the IDB's world population projections in the course of generating uncertainty information that we compare to the uncertainty information developed by the United Nations for its world population projections (Alkema *et al.* 2015). The approach we suggest employs the ARIMA (Auto-Regressive Integrated Moving Average) Time Series method in conjunction with work by Espenshade and Tayman (1982) whereby we can translate the uncertainty information found in the ARIMA method's forecast to the population forecast provided by the CCM approach.

In the remainder of the paper, we describe the ARIMA model along with the source data we use in this paper and then show how to translate its uncertainty information to a CCM forecast. Following the evaluation of the uncertainty information generated for the world as a whole, we then generate an ARIMA forecast to 2050 for Estonia and translate its 95% confidence intervals to national population projections found at the International Date Base (IDB) of the U.S. Census Bureau (2020). Following Swanson and Tayman (2014), who argue that 95% confidence intervals may be too wide to be useful, we also provide 66% confidence intervals.

We do not describe the CCM approach in any detail because it is so widely known and used (Baker et al., 2017; George et al., 2004; Smith, Tayman, and Swanson, 2013). We also stress that the forecast for Estonia is heuristic. That is, we have selected Estonia as a case study of the method and our work is not intended to compete with official (and other) forecasts of Estonia's population (Maamgägi, 2007; Statistics Estonia, 2024).

### The ARIMA Model

At its heart, the ARIMA (Auto-Regressive Integrated Moving Average) time series model is a regression-based projection method. It was popularized by Box and Jenkins (1976) and has been used in the analysis and projection of business, economic, and demographic variables. Examples of its use in demographic forecasting include McNown et al. (1995); Pflaumer (1992); Swanson (2019); Tayman, Smith, and Lin (2007); and Zakria and Muhammad (2009).

As discussed by Smith, Tayman, and Swanson (2001: 172-176), an ARIMA model attempts to uncover the stochastic processes that generate a historical data series. The mechanism of this stochastic process is described—based on the patterns observed in the data series—and that mechanism forms the basis for developing projections. Up to three processes can describe the stochastic mechanism: autoregression, differencing, and moving average. The autoregressive process has a memory in the sense that it is based on the correlation of each value of a variable with all preceding values. The impact of earlier values is assumed to diminish exponentially over time. The number of preceding values explicitly incorporated into the model determines its "order." For example, in a first-order autoregressive process, the current value is explicitly a function only of the immediately preceding value. However, the immediately preceding value is also a function of the one before it, which is a function of the one before it, and so forth. Consequently, all preceding values influence current values, albeit with a declining impact. In a second-order autoregressive process, the current value is explicitly a function of the two immediately preceding values; again, all preceding values have an indirect impact.

The differencing process is used to create a stationary time series (i.e., one with constant differences over time). A stationary time series is very important for the construction of ARIMA models. When a time series is non-stationary, it can often be converted into a stationary time series by calculating differences between values. First differences are usually sufficient, but second differences are occasionally required (i.e., differences between differences). Logarithmic and square root transformations can also be used to convert non-stationary to stationary time series.

The moving average represents a "shock" to the system or an event that has a substantial but short-lived impact on the time series pattern. This impact has a limited duration, and then time series trends return to normal. The order of the moving average process defines the number of time periods affected by the shock. The most general ARIMA model is usually written as ARIMA (p, d, q), where p is the order of the autoregression, d is the degree of differencing, and q is the order of the moving average. (ARIMA models based on time intervals of less than one year may also require a seasonal component.) The first and most subjective step in developing an ARIMA model is to identify the values of p, d, and q. The d-value must be determined first because a stationary series is required to properly identify the autoregressive and moving average processes. The value of d is the number of times one has to difference the series to achieve stationarity (usually 0 or 1, but occasionally 2). The p- and q-values are also relatively small (0, 1, or—at most—2). The patterns of the autocorrelation (ACF) and partial autocorrelation functions (PACF) are used to find the correct values for p and q. For example, a first-order autoregressive model [ARIMA (1, 0, 0)] is characterized by an ACF that declines exponentially and quickly and a PACF with a significant value only at lag 1. Once p, d, and q are determined, maximum likelihood procedures are used to estimate the parameters of the ARIMA model. The final step in the estimation process is model diagnosis. An adequate ARIMA model will have random residuals, no significant values in the ACF, and the smallest possible values for p, d, or q. After a successful diagnosis is completed, the ARIMA model is ready to use.

As alluded to earlier, underlying the Espenshade-Tayman method is the idea that there is a sample taken from a population of interest. In this case, the ARIMA results represent the sample and the CCM forecasts represent the population. This interpretation is derived from the idea of a "superpopulation" (Hartley and Sielken, 1975; Sampath, 2005; Swanson and Tayman (2012: 32-33). This concept can be traced at least back to Deming and Stephan (1941) who observed that even a complete census, for scientific generalizations, describes a population that is but one of the infinity of populations that will result by chance from the same underlying social and economic cause systems. It is a theoretical concept that we use to simplify the application of statistical uncertainty to a population forecast that is considered a statistical model in this context. This approach is conceptually and mathematically different from the classical frequentist theory of finite population sampling (Hartley and Sielken (1975), but as pointed out by Ding, Li, and Miratrix (2017), in practical terms, these two approaches result in identical variance estimators. As such, we believe that our approach is on solid statistical ground. Before moving on, we note that the use of the Espenshade-Tayman method (1982) here is not new. In addition to being employed by Espenshade and Tayman (1982), it has been used by Swanson (1989) and Roe, Swanson, and Carlson (1992) in demographic applications.

## 2. Methods and Data

By employing the ARIMA (Auto-Regressive Integrated Moving Average) Time Series method in conjunction with work by Espenshade and Tayman (1982), we can translate the uncertainty information found in the ARIMA method's forecast to the population forecast provided by the CCM approach. We note that the patterns of the autocorrelation (ACF) and partial autocorrelation functions (PACF) were used to find the correct values for p and q (Brockwell and Davis, 2016: Chapter 3). The ARIMA models we describe for the world as a whole and for Estonia had random residuals and the smallest possible values for p, d, or q, as determined by the Ljung-Box test (Ljung and Box, 1979). We chose an "adequate" ARIMA model using these criteria for both the world as a whole and Estonia. We note that there may be other versions that also are "adequate" and that further refinement of the selection process can be done (e.g., using the augmented Dickey-Fuller test (Dickey and Fuller, 1979) to identify the amount of differencing required to achieve a stationary time series). Because our aim here is heuristic not definitive, we did not pursue further refinement of the ARIMA model we present beyond determining it to be adequate (See the Appendix). We note, however, that if one intends to use our approach, the selection of an ARIMA model needs to be consistent with guidelines described here and elsewhere (e.g., Box and Jenkins, 1976; Brockwell and Davis, 2016; Hyndman and Athanasopoulos, 2021; Smith, Tayman, and Swanson, 2013; Swanson and Tayman, 2024).

Before describing the new method, we first clarify our use of the term "confidence interval" regarding forecast uncertainty. It is more common to use the term "forecast interval" or "prediction interval" in the context of forecasting because a "confidence interval," strictly speaking, applies to a sample (Swanson & Tayman, 2014: 204). However, underlying the approach we describe herein is the concept of a "superpopulation," which, as discussed later, represents a population that is but one sample of the infinity of populations that will result by chance from the same underlying social and economic cause systems (Deming & Stephan, 1941). Viewing a forecast as a sample leads us to use the term "confidence interval" rather than a forecast or prediction interval.

### 3.     Probabilistic Estonian Population Forecast

We use annual world historical data of total population and land area in square meters to compute population density annually from 1950 to 2020 found at the IDB site to implement the ARIMA (Box-Jenkins) model found in the NCSS statistical package (NCSS, 2024) and launch from the annual world forecasts found at the same site for 2021-2060. We use "density" because the Espenshade-Tayman (1982) method for translating uncertainty information does so from an estimated "rate," which in this case is the "rate" of population density. Thus, the 95% confidence intervals generated by the ARIMA world "density" forecasts are translated to the CCM-based world population forecast. Other denominators could be used in developing such a "rate, such as the ratio of the population to housing units. However, using the land area as the denominator provides a virtually constant denominator over time, thereby reducing the effort in assembling the "rate" data. It also serves as a stabilizing element regarding the use of ARIMA in that it dampens the effect of short-term population fluctuations more effectively than, say, housing units, which also can fluctuate over time and not always in concert with population fluctuations. As should be obvious, the data assembled to develop the ARIMA density forecast should encompass the base data used to develop the population projection in terms of the total population numbers. The case study we present meets this condition in that the annual ARIMA model covers the period from 1950 to 2020 and the  IDB population projection of Estonia is launched from 2020 and employs earlier data to develop the base data used in the launch.

 Here is an example of this process using the 2050 Estonian forecast data at the IDB site.

Let P = forecasted Estonian population (at time t) obtained from the forecast,

Let D = forecasted Estonian population density obtained from ARIMA at time t, and

Let A = land area of Estonia (42,388 square kilometers).

The 2050 ARIMA density forecast shows 14.5436, 25.2596, and 35.3582 persons per square kilometer, respectively, for the land area of the world as a whole (95% Lower Limit of forecasted D, forecasted D, and 95% Upper Limit of forecasted D, respectively).

The relative widths of the Upper and Lower Limits are -0.4242 and 0.4242, respectively.

The 2050 Estonia forecast found at IDB is 970,580

Multiplying 970,580 by -.4242 and adding this product to 970,580 yields 558,826, the 95% Lower Limit, and adding the product 979,580× ,4224 to 9970580 yields 1,382,334, the 95% Upper Limit of the 2050 Estonian population forecast found at IDB.

Putting it all together, we can state that we are 95% certain that the 2050 Estonian forecast found at IDB is between 558,826 and 1,382,334

(TABLES 1.A THROUGH 1.D ABOUT HERE)

The steps in going from Tables 1.A through 1.D are those described earlier. Once at Table 1.D, one can see that as forecast moves from 2030 (1,138,017 with lower and upper 95% confidence interval bounds of 1,043,514 and 1.232,524, respectively) to 2050, the interval widths increase over time, with a forecast of 970,580 and lower and upper 95% confidence interval bounds of 558,826 and 1.382,334, respectively. Like the projections described by Statistics Estonia (2024), the probabilistic forecasts based on the IDB projections show a decline in population.

## 4. Discussion

As is the case with the Bayesian approach described by Alkema *et al*. (2015), the new approach we propose can be linked directly to the CCM method (as well as forecasts produced by other methods such as the Cohort Change Ratio (CCR) approach, which is algebraically equivalent to the CCM approach, but requires less input (Baker et al., 2017)). Unlike the approach found in Swanson and Beck (1994), neither the CCM nor the CCR approach is inherently conjoined with a method for generating statistical uncertainty. Thus, we believe this linkage represents a step forward on the path to generating probabilistic forecasts based on the fundamental population equation. Notably, the ARIMA method is widely available in the software packages generally used by demographers.

The approach we propose does not produce the uncertainty intervals by age and gender, births, death, and migration as does the Bayes CCM approach described by Alkema *et al*. (2015), Yu *et al*. (2023: 934) and the CCR approach discussed by Swanson and Tayman (2014). However, neither the Bayes CCM nor our approach take into account uncertainty in the input data themselves. However, as Yu et al. (2023: 934) implied, these are not likely to be among the most important sources of uncertainty where population forecasts are routinely produced.

Regarding our approach not providing uncertainty intervals by age and gender, Deming's (1950: 127-134) "error propagation" was used to translate uncertainty in age group intervals found in the regression-based CCR forecasts reported by Swanson and Tayman (2014) to the total populations in question. In different forms, "error propagation" has been used by Alho and Spencer (2005), Espenshade and Tayman (1982), and Hansen, Hurwitz, and Madow (1953), among others. It may be possible to reverse-engineer error propagation and develop uncertainty measures by age and gender using our approach. The validity of this could be explored to determine if it is viable. As an approximation, one could generate age uncertainty intervals by controlling "low" and "high" numbers in a given forecast series to their corresponding 95% lower and upper limits, respectively, found using our proposed approach.

It is important to keep in mind that we used information found at the U.S. Census Bureau's International Data Base to generate the probabilistic population forecast for Estonia, which was a convenient source because all of the information we needed for both the world as a whole and Estonia was in a common source. The projections generated by the UN for the world as a whole are similar but not the same as those found at the IDB site. Similarly, population projections generated in Estonia (see, e.g., Maamgägi, 2007; Statistics Estonia, 2024) vary from those found at the IDB site. In this regard, we note again that our purpose here is heuristic rather than definitive. With these points in mind, however, it is clear that like the probabilistic world population forecasts the uncertainty boundaries for the Estonian probabilistic forecasts become wider over time, an important feature consistent with knowledge about forecast uncertainty (Swanson, Tayman, and Cline, 2024). In addition, the uncertainty boundaries at each year are wider than those for the corresponding world population forecasts, another important feature consistent with knowledge about forecast uncertainty in that there is more uncertainty found in forecasts of a small population than in a large population, ceteris paribus (Swanson, Tayman, and Cline, 2024).

We (Swanson and Tayman, 2025) applied this approach to world population projections and compared the results to the Bayesian-based probabilistic world forecast produced by the United Nations (Alkema et al., 2015; Raftery, Alkema, and Gerland, 2014, United Nations, 2022, 2024), which we found to be similar but with more uncertainty than found in the latter. We (Swanson and Tayman, 2024) also used historical data produced by the Forecasting Division of the Office of Financial Management (Washington, 2024) to implement the ARIMA model, which we then applied to the medium series of the "Growth Management Act" projections produced by the Office of Financial Management (Washington, 2022).

We examined the range of uncertainty in the county forecasts by analyzing half-widths and compared the half-widths of their ARIMA-based intervals to half-widths produced using a Bayesian approach by Yu et al. (2023, who discussed their results for the state as a whole and to three counties, Ferry, King, and Whitman. The ARIMA-based half-widths for Ferry county (a county with a very small population, approximately 7,500 currently) were found to be wider than those for the Bayes CCM approach at each of the three horizon lengths, 10, 20, and 30 years. For King county (the county with the highest current population of Washington's counties, approximately 2.7 million), the ARIMA-based half-widths were slightly narrower at 10 years than those found for the Bayes CCM approach and substantially narrower at 20 and 30 years. For Whitman County (which has a current population of approximately 48,000, of which 27,000 or so are students at Washington State University), the Bayes-based CCM produced narrower widths at each of the three horizon lengths. However, we noted that Yu et al. (2023: 921-922) held the age groups associated with college attendance constant in counties such as Whitman where these populations have a large impact on the age structure of the county as a whole. Considering Washington State as a whole (with a population of approximately 7.9 million), we found that the half-widths generated by the Bayes CCM were narrower for the 10- and 20-year horizon than the ARIMA-based half-widths, while the latter produced a narrower half-width for the 30-year horizon length. Importantly, for all three counties and the state as a whole, the ARIMA-based half-widths increased over time in a manner consistent with the Bayes CCM half-widths. We (Swanson and Tayman, 2024) concluded that the ARIMA-based approach produces uncertainty measures for county population forecasts that are not dissimilar to those produced by the Bayes CCM approach and that neither produced intervals so wide as to be useless, a point brought up by Swanson and Tayman (2016) in an earlier examination of forecast uncertainty. In regard to this point, however, Swanson and Tayman (2014) argued that 95% CIs are likely too wide to be of use and recommended that 66% CIs be considered. Following this recommendation, the following table (1.F) shows the results for Estonia using 66% CIs, which in our opinion are preferable to the 95% CIs shown earlier.

(TABLE 1.F ABOUT HERE)

## 5. Conclusion

In closing, we argue that the approach we propose and have described in this paper is well-suited for generating not only probabilistic world and national forecasts, but also subnational population forecasts where these forecasts are routinely produced. Because it can be applied to both the CCM and the CCR approaches, our method for producing forecast uncertainty information provides a path to a reasonable level of forecast accuracy as identified by Swanson et al. (2023). It also has the potential to optimize forecast utility. None of this is meant to imply that forecast uncertainty measures derived from ARIMA models using the Espenshade-Tayman method are more "accurate" than those generated from a Bayesian method. It is simply an alternative.

.

**Declarations and Acknowledgements**

The data underlying this paper are secondary and available from the corresponding author.

This research was not funded by any agency.

The authors have no conflicting interests regarding this paper.
An ethics approval statement is not applicable because the data used are secondary, and no human subjects review was required.

A patient consent statement is not applicable.

A clinical trial registration statement is not applicable because no human subjects were involved.

A statement regarding permission to reproduce material from other sources is not applicable because no excerpts from copyrighted works owned by third parties are included.

We thank Luule Sakkeus and her colleagues at the Estonian Institute for Population Research for comments and suggestions on an early version of this paper and also for their comments, the participants in Session 6.3, Innovations in Demographic Forecasting and Projection, at the 2025 Nordic Demographic Symposium

| | Table 1.A ESTONIA POPULATION FORECAST | | |
|---|---|---|---|
| NATION | **2030** | **2040** | **2050** |
| **ESTONIA** | 1,138,017 | 1,052,590 | 970,580 |

**Table 1.B ARIMA DENSITY FORECAST (ABSOLUTE)**

| 2030 | | | 2040 | | | 2050 | | |
|---|---|---|---|---|---|---|---|---|
| FORECAST | LL95% | UL95% | FORECAST | LL95% | UL95% | FORECAST | LL95% | UL95% |
| 27.1587 | 24.9034 | 29.4141 | 25.9782 | 19.6344 | 32.322 | 25.2596 | 14.5436 | 35.9756 |

**Table 1.C ARIMA DENSITY FORECAST (RELATIVE)**

| 2030 | | 2040 | | 2050 | |
|---|---|---|---|---|---|
| LL95% | ULP5% | LL95% | UL95% | LL95% | UL95% |
| -0.08304153 | 0.083045212 | -0.244197058 | 0.244197058 | -0.424234746 | 0.424234746 |

| TABLE 1.D | 2030, 2040, AND 2050 NATIONAL POPULATION FORECASTS AND THEIR 95% UNCERTAINTY INTERVALS | | | | | | | | |
|---|---|---|---|---|---|---|---|---|---|
| | 2030 | | | 2040 | | | 2050 | | |
| NATION | FORECAST | LL95% | UL95% | FORECAST | LL95% | UL95% | FORECAST | LL95% | UL95% |
| **ESTONIA** | 1,138,017 | 1,043,514 | 1,232,524 | 1,052,590 | 795,551 | 1,309,629 | 970,580 | 558,826 | 1,382,334 |

| TABLE 1.F | 2030, 2040, AND 2050 NATIONAL POPULATION FORECASTS AND THEIR 66% UNCERTAINTY INTERVALS | | | | | | | | |
|---|---|---|---|---|---|---|---|---|---|
| | 2030 | | | 2040 | | | 2050 | | |
| NATION | FORECAST | LL66% | UL66% | FORECAST | LL66% | UL66% | FORECAST | LL66% | UL66% |
| **ESTONIA** | 1,138,017 | 1,090,221 | 1,185,813 | 1,052,590 | 922,069 | 1,183,111 | 970,580 | 760,935 | 1,180,225 |

**Appendix. ARIMA Report**

Dataset          C:\...\ESTONIA\ESTONIA AREA POP DENSITY 1950-2023.NCSS
Variable         DENSITY-TREND

## Minimization Phase Section

| Itn No. | Error Sum of Squares | Lambda | AR(1) |
|---|---|---|---|
| 0 | 5.484014 | 0.1 | 0.1 |
| 1 | 0.6494514 | 0.1 | 0.8593766 |
| 2 | 0.5727309 | 0.04 | 0.9553774 |
| 3 | 0.5724399 | 0.016 | 0.9612668 |
| 4 | 0.5724401 | 0.0064 | 0.9616405 |
| 5 | 0.57244 | 0.064 | 0.9616203 |
| 6 | 0.5724397 | 0.64 | 0.9614962 |

Normal convergence.

## Model Description Section

| | |
|---|---|
| Series | DENSITY-TREND |
| Model | Regular(1,1,0)   Seasonal(No seasonal parameters) |
| Trend Equation | (30.80116)+(0.02418899)x(date) |
| | |
| Observations | 74 |
| Missing Values | None |
| Iterations | 6 |
| Pseudo R-Squared | 99.915050 |
| Residual Sum of Squares | 0.5724397 |
| Mean Square Error | 0.007950552 |
| Root Mean Square | 0.08916587 |

## Model Estimation Section

| Parameter Name | Parameter Estimate | Standard Error | T-Value | Prob Level |
|---|---|---|---|---|
| AR(1) | 0.9614962 | 0.03059133 | 31.4303 | 0.000000 |

## Forecast and Data Plot

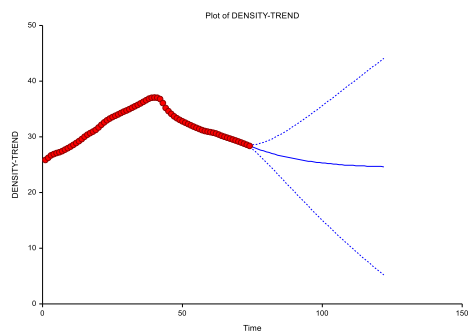

**Autocorrelations of Residuals of DENSITY-TREND**

| Lag | Correlation | Lag | Correlation | Lag | Correlation | Lag | Correlation |
|---|---|---|---|---|---|---|---|
| 1 | 0.400167 | 13 | 0.090163 | 25 | 0.063520 | 37 | 0.086111 |
| 2 | -0.075988 | 14 | 0.025157 | 26 | 0.077727 | 38 | 0.149888 |
| 3 | -0.016163 | 15 | -0.128197 | 27 | 0.028183 | 39 | 0.157346 |
| 4 | -0.024545 | 16 | -0.114217 | 28 | -0.072754 | 40 | -0.009002 |
| 5 | -0.142687 | 17 | 0.033592 | 29 | -0.084807 | 41 | -0.146874 |
| 6 | -0.141474 | 18 | 0.055798 | 30 | -0.010600 | 42 | -0.043296 |
| 7 | -0.075700 | 19 | 0.055508 | 31 | -0.038922 | 43 | -0.023557 |
| 8 | -0.085602 | 20 | 0.101835 | 32 | -0.081575 | 44 | 0.000727 |
| 9 | -0.043228 | 21 | 0.010672 | 33 | -0.035905 | 45 | 0.023018 |
| 10 | -0.011662 | 22 | -0.130101 | 34 | -0.007455 | 46 | 0.025295 |
| 11 | -0.040070 | 23 | -0.169485 | 35 | -0.021691 | 47 | 0.017776 |
| 12 | 0.002167 | 24 | -0.054299 | 36 | -0.030857 | 48 | 0.007455 |

Significant if |Correlation|> 0.232495

**Autocorrelation Plot Section**

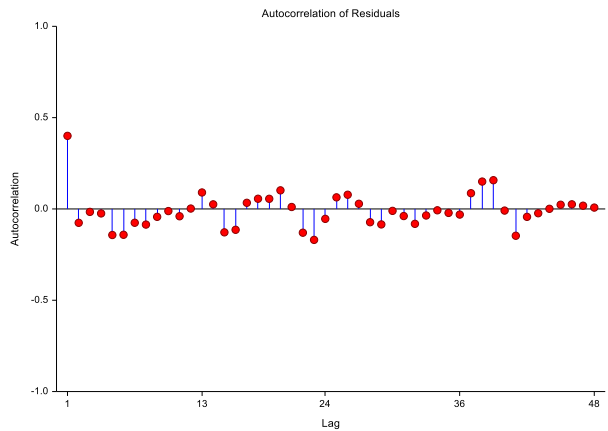